# Modeling Inequality in Complex Networks of Strategic Agents using Iterative Game-Theoretic Transactions

**Mayank Kejriwal**[1,*], **Yuesheng Luo**[1]

[1] *Information Sciences Institute, Viterbi School of Engineering, University of Southern California, Marina del Rey, CA, United States of America*

Correspondence*:
Mayank Kejriwal
kejriwal@isi.edu

## ABSTRACT

Transactions are an important aspect of human social life, and represent dynamic flow of information, intangible values, such as trust, as well as monetary and social capital. Although much research has been conducted on the nature of transactions in fields ranging from the social sciences to game theory, the systemic effects of different types of agents transacting in real-world social networks (often following a scale-free distribution) are not fully understood. A particular systemic measure that has not received adequate attention in the complex networks and game theory communities, is the Gini Coefficient, which is widely used in economics to quantify and understand wealth inequality. In part, the problem is a lack of experimentation using a replicable algorithm and publicly available data. Motivated by this problem, this article proposes a model and simulation algorithm, based on game theory, for quantifying the evolution of inequality in complex networks of strategic agents. Our results shed light on several complex drivers of inequality, even in simple, abstract settings, and exhibit consistency across networks with different origins and descriptions.

## 1 INTRODUCTION

With conceptual and methodological advances in both network science [1], and computational social science [2], it has become possible to study complex research questions by modeling and simulating the evolution of dynamic social systems [3], [4]. Research on social networks, within the computational sciences alone, now spans over two decades of research, with recent focus on higher-order and 'multiplex' networks [5], [6].

Game theory has also played a prominent role in studies of networks over the last decade. Historically, game theory largely found its applications in economics and studies of decision-making under conditions of uncertainty [7], [8]. Much more recently, a growing body of work has explored the utility of game-theoretic techniques in modeling agent-based interactions in graphs and networks [9], [10]. This work is motivated by the fact that, although networks have a long history in modeling and simulating human social systems, such as work in network growth models that aim to model the dynamic aspects of such systems (including how the scale-free structure often observed in such networks comes to be), there is less work on using the network to model a system of *dynamic transactions*.

A transaction here does not have to be a monetary transaction, since it can involve intangible goods, such as goodwill, social capital, trust, information and other 'goods' on which there has been much exposition in the social sciences [11], [12]. We adopt a dictionary definition of the word here, with Merriam-Webster defining the word 'transaction' as both (emphases ours) "an *exchange* or *transfer* of goods, services, or funds" and as "a communicative action or activity involving two parties or things that *reciprocally affect* or *influence* each other"

Arguably, a significant fraction of interactions in everyday social life is transactional, rather than growth-based where *new* nodes or actors form a social acquaintance with us. Intuitively, the majority of our interactions tend to be largely limited to people we are already connected to, whether friends, family, neighbors and colleagues. Yet, research at the intersection of game theory and networks has tended to focus more on network growth, including game-theoretic explanations for models (sometimes, but not always, relying on simulations) such as preferential attachment [13].

Transactions also play an important role in the evolution of macroscopic properties such as *inequality*. Questions of wealth and income inequality play a major role in economics, with entire books written on the subject over the decades [14], [15], [16]. Worryingly, inequality has grown to alarming levels in recent times, with some directly blaming monetary policies (instituted by central banks, such as the US Federal Reserve) for fueling rises in asset prices, which disproportionately benefit the (already relatively) wealthy, and higher-income individuals [17].

Without denying the complexities of the causes and effects of inequality in large-scale economies, we hypothesize that more modest claims can be made by conducting controlled simulations on reasonably large-scale, complex networks of *transacting agents.* As we subsequently discuss in *Model*, such simulations assume an initial 'social' network of transacting agents where every agent starts off with the same amount of 'capital' (assumed to be denominated in dollars, for the sake of simplicity). The system is therefore perfectly equal, as quantified by a *Gini Coefficient* (for which we provide a formula and background in subsequent sections) [18] of 0 for the entire system.

In such a setup, which largely draws on empirical simulations and measurements rather than *a priori* theories, the types, mixtures and assignments of agents to nodes in the network depend on the experimental controls, and the research issue to be investigated. Another important parameter that is largely absent from work involving social networks (in general) but that occupies a central position in our experiments and model, is that of an *external authority*. In contrast with network science, such an authority is usually implicitly present in famous models of game theory (especially involving games that are not purely zero-sum), such as the Prisoner's Dilemma [19]. In networks with financial entities, such authorities typically represent government entities, and examples of transactions between such external authorities and the agents in the network include asset purchases [20], bailouts [21], or even non-monetary but costly interventions such as securities investigations and anti-trust regulation [22], [23].

With these motivations in mind, our goal in this article is to conduct a set of systematic empirical studies that draw on simple, but important, principles at the intersection of game theory and agent-based transactions, and network structure, to investigate how the inequality of the system evolves over time, and how strong the effects of even a simple external authority can potentially be. Specific contributions made in this article are as follows:

1. We present a model and novel algorithm for applying the Prisoner's Dilemma framework on collections of decision-making agents that exhibit some key properties also found in the real world: bounded information and rationality [24], influence of neighbors [25], incentive structures presented by external

   authorities (such as government regulators) [26], and availability of limited resources. Our model is simple in that it uses real-world networks and the classic payoff matrix of the Prisoner's Dilemma [19], in conjunction with a simple model for the external authority.

2. Our model measures a systemic property (inequality), using the Gini Coefficient measure [18], and its change over time as we simulate our algorithm on networks representing sets of interconnected agents.

3. We propose two novel sets of experiments (with over ten experimental settings and controls investigated in total), which consider the effects of independent variables such as network structure, external authority, and the proportions of agents implementing different strategies, on the Gini Coefficient of the system.

4. Using three real-world and publicly available networks, including a social network, an academic collaboration network, and a cryptocurrency transaction network, we conduct a series of experiments by simulating strategic decision-making over many iterations, and report detailed results on how the Gini Coefficient evolves over time.

5. Our experimental results show that the presence of the external authority can have significant (and often, unwanted) effects on inequality, especially when the external authority has unlimited reserves and capacity. To our knowledge, such an effect has not been empirically shown before on real-world networks, with consistency observed across the networks (despite their different origins and properties).

The remainder of this article is structured as follows. We begin with a discussion of related work in Section 2, followed by a description of our model and algorithm in Section 3. We then discuss our experimental setup, and materials and methods, in Section 4. Results follow in Section 5, with a summary of core findings and additional discussion in Section 6. The article concludes with some guidance on future work in Section 7.

## 2 RELATED WORK

There is a considerable amount of recent research on using game theory to explain popular network formation models such as preferential attachment. An excellent example is [13], wherein the authors show that preferential attachment is a unique and universal Nash equilibrium when the network growth is modeled as a wealth recommendation game. They also cite a broad body of research by other authors who have attempted to use games (in the game theoretic sense) in a similar fashion. Some selected examples include [9], [10], [27], [28], [29]. Some of this work, such as [28], is based on the notion of 'network reputation games' where participants can rate other participants by forming links. This kind of network was originally useful for studying hyperlink placements and PageRank optimization for the Web, but has since found many uses in other online networks (such as social networks) where reputation also matters greatly [30], [31]. Another good example, especially relevant to economics, is the work in [32], which studies the emergence of cooperation in a public goods game with a reputation-updating timescale. Other recent work that seeks to investigate cooperation and its emergence in a similar manner include [33] and [34], with the latter also considering public goods games.

Similarly, [35] considers how *adaptive* reputation promotes trust in social networks. The authors in that paper conduct Monte Carlo simulations to determine critical values of the degree of rationality in social networks, as well as showing how a reputation threshold can be associated with higher levels of trust and 'social wealth'. Similar other papers include [36], [37]. In yet other work of a more social science nature, non network-based models of 'social wealth' or social capital have also been investigated, examples including [38] and [39], with the former presenting an equational model, and the latter, a more descriptive model.

There are close connections between these works, and other issues such as information dissemination [40], [41], interactive sensing [42], [43], and optimization [44], [45]. Other than network reputation games, much of the related work considers game theory as an underlying explanation for network formation and growth models, as noted at the beginning of this section.

Our work is complementary to these lines of research, but also distinct in that we are not aiming to explain how the network was actually formed, but how inequality can take hold in a network (even when all agents start off on an even footing) due to various effects, such as the collection, proportions of, and distribution function underlying the strategies adopted by the agents assigned to the nodes in the network, and the presence of an external authority. A commonality shared by network growth models and the model proposed herein is that they are both dynamic. While the network structure, including numbers of nodes and edges are changing over time in the related literature, our model considers the network to be mature (i.e., fully formed and unchanging in structure) but transactions simulated on the network are dynamic, leading to fluctuating inequality over time.

We draw heavily upon the Gini Coefficient measure in this article for measuring and studying inequality in the system. A formula and definition is provided subsequently when we describe our model, but herein, we note that this measure is well established for studying economic inequality [18], [46]. National-level Gini Coefficient measures are typically available over time for most nations and have been studied by scholars [47], with detailed studies available for recently developed economies such as China, especially from the perspective of trend analysis [48]. However, to our knowledge, this is the first work using Gini Coefficient to understand inequality in a complex system, represented as a network, with transactions mediated using a game such as Prisoner's Dilemma.

Finally, we note that within agent-based social sciences (especially, economics), game theory has a rich history and scholarly tradition [7], [8], extending to Von Neumann's original treatment more than a half-century ago (a recently re-published edition of which is [49]). The Prisoner's Dilemma model has been particularly influential [19], and its influence only grew in the computational era when Axelrod conducted a competition based on the Iterated Prisoner's Dilemma (IPD) [50], [51]. The tournament continues to be a popular framework even today for studying agent strategies for IPD [52], [53], including in the presence of noise [54]. Although the model we consider in this article is similar in structure (i.e., the agents in our networks are also simulated to play Prisoner's Dilemma repeatedly), it is not the same as IPD, since the agent against which a given agent plays is determined by the network structure, and not necessarily fixed (even strategically) over a sequence of games. Furthermore, we do not consider the issue of strategic optimization in this article, since good solutions to that (including the 'tit-for-tat' strategy that we consider as a candidate decision-making agent in this article) have been fairly evident over a series of papers in the IPD literature. Instead, we are seeking to understand the how inequality evolves in complex networks in various experimental settings that are subsequently discussed.

## 3 MODEL

The model of transacting agents is fundamentally defined in this article as an undirected graph $G = (V, E)$, where V is the set of nodes or vertices, and E is the set of undirected edges. Based on prior discussion, nodes should be thought of as individuals or users who are aiming to transact with one another given the

|   | B stays silent or 'cooperates' | B betrays or 'defects' |
|---|---|---|
| A stays silent or 'cooperates' | +1 / +1 | +3 / -3 |
| A betrays or 'defects' | -3 / +3 | -2 / -2 |

**Figure 1.** The Prisoner's Dilemma payoff matrix that is used for simulating node-pair transactions in this article.

Prisoner's Dilemma payoff matrix in Figure 1. The transactions are constrained by the structure of the network i.e., a node cannot transact with another node unless it is directly linked to it via an edge.

Real-world social and transactional networks (such as a scientific collaboration network and the Bitcoin network we use in this article) are not usually random or even Gaussian in their degree distributions, but tend to be scale-free [1]. To account for this skewness and allow each node to have a fair opportunity to transact, we allow each node a minimum of one transaction in each iteration. By iteration, we mean a traversal of all nodes in an order that is decided randomly before we begin simulating the model. Although the ordering is random, it stays fixed throughout the duration of the entire experiment (both the first iteration as well as other iterations that follow it). The specific steps in the simulation are enumerated in Algorithm 1.

In Algorithm 1, before beginning the first iteration, we assign each node in $G$ to exactly one of four classic decision-making agents in the game theory literature: *Cooperator, Defector, Tit-for-Tat*, and *Random*. Agent decision making behavior for each of these (mnemonically named) models is discussed in the next section. The manner in which a node is assigned to an agent model depends on the specific experiment, as detailed in Section 4 wherein we describe experimental methodology.

Each node is parameterized by an amount called the *current balance*. The current balance is the amount of capital that the node currently (i.e., in this iteration) possesses, and can potentially transact with. The current balance can have the interpretation of dollar amounts, bitcoin or even social capital [11], such as likes and dislikes in a social network. Nodes cannot possess negative current balances. Each node also starts out with an initial current balance $i$, which is a parameter that is an input to the algorithm. In our experiments, we always set $i$ to 100. Every node in the network always starts out with the same initial current balance; hence, there is perfect equality prior to the first iteration.

In line 2 of Algorithm 1, a random order is decided in which the nodes will be traversed, which stays fixed for the remainder of the algorithm's execution. This order is captured by the list $V'$ of nodes, rather than the set $V$ of nodes that has no order and is the original input in the algorithm as part of the graph $G$. The list $L$, which will be the output of the algorithm, is initialized as empty in line 3. The size of this list will exactly equal the number of iterations (upwardly bound by $N$, which is also an input parameter) actually executed by the algorithm. As we shortly describe, this number could potentially be less than $N$, if convergence is obtained before the $N^{th}$ iteration. The $k^{th}$ item in the list will correspond to the Gini Coefficient of the entire system, for which a formula is shortly provided, measured after the conclusion of the $k^{th}$ iteration. The list $L$ allows us to measure how inequality is evolving in the system over time. In the loop in line 4, we assign two variables to each node to maintain a limited memory (which will prove useful for the purposes of checking if the system has converged).

In line 6, the simulation properly begins with the first iteration. In each iteration, we always traverse the (randomly ordered, but fixed) list $V'$ of nodes, in order, from beginning to end. Note that the list is only guaranteed to be fixed for that algorithmic run i.e., a different random seed could 're-order' the list once a different experiment is conducted or if the same experiment is conducted again independently. Empirical, we did not find the ordering of the list to appreciably affect the statistical outcome of the simulation for a fixed set of inputs.

As shown in the algorithm, per iteration, we first randomly sample a neighbor ('opponent') of the 'current node' (determined by the iteration's current position in the list). The agent models of the two nodes (the current node and its opponent) now make decisions. Based on the decision, the nodes exchange a portion of their balances if one of them betrays and the other one stays silent. We use the terminology of the payoff matrix in Figure 1. This is the classic zero-sum situation. However, if both of them stay silent or betray, the Prisoner's Dilemma is not designed to be zero-sum. In keeping with this intuition, when both nodes make the same decision, the model is triggered to facilitate an exchange between each node and the *bank*, which is an external entity that serves as a global reservoir or store of value. The bank does not have to be a literal 'bank', but can be an escrow authority, or even governmental, in nature. Note that the Prisoner's Dilemma implicitly assumes an external authority in its usual description (the 'law enforcement' authorities whose goal is to extract a guilty plea from one or more of the two prisoners). In our model, specifically, when both nodes choose stay silent, each receives one unit from the bank, while if they both betray, they have to surrender two units (each) to the bank. The bank has unlimited capacity (i.e., there is no set limit on the amount of capital it can be a reservoir for), but receives an initial balance of its own, depending on the experiment. We consider three initial balances for the bank (0, 10,000 and infinite). Except in the infinite case, the bank itself has finite capital to start with, and by definition, the total amount of capital in the entire system (the balance of all nodes, and the bank) always exactly equals the sum of the initial bank balance and $|V|.i$. The bank's initial balance can have a significant impact on the evolution of inequality in the model, as we empirically illustrate in Section 5.

There are some corner cases that must be borne in mind with respect to the transactions (between a pair of nodes), represented by the many 'if-then-else' statements within the iteration. First, if either node has zero current balance, the 'game' or transaction is skipped. That is, we move on to the next node in the list. Note that we do not 're-sample' another neighbor if a transaction does not succeed with the opponent first sampled, as it would lead to an outcome subject to unwanted selection bias. Second, in the event that a node has non-zero balance $p$, but the balance is less than the eventual (absolute) transaction amount $t$ (e.g., $t = 2$ if both nodes end up betraying) the node with $p < t$ would only have to pay out its full current balance. It could potentially *receive t*, however. For example, suppose that node A has a current balance of 2, node B has a current balance of 4, and A betrays while B stays silent. B would then have to forsake 3 of its 4 units to A. If it happened the other way around, A would yield its 2 remaining units to B, which would

---

**Algorithm 1** Simulating Prisoner's Dilemma transactions on network of decision-making agents

**Input :**
- Network model (undirected, unweighted graph): $G = (V, E)$
- Number of iterations : $N$
- Prisoner's Dilemma payoff matrix : $P$
- Bank initial balance: $b$ # *must always be* $\geq 0$; *set to NULL if is_bank_infinite is set to True*
- Bank infinity flag (Boolean): $is\_bank\_infinite$ # *if this flag is set to True, b will not be used*
- Initial current balance: $i$

**Output :**
- A list $L$ with (at most) $N$ elements, each representing a Gini Coefficient

**Steps :**
1. Assign each node in $V$ to an agent model in *[Cooperator, Defector, Random, Tit-for-Tat]* (based on the experimental setup);
2. $V' := list(V)$ that is ordered using a random seed;
3. Initialize $L$ as empty list;
4. **for** each node $v$ in $V'$ **do**
    Set $v.current\_balance := -1$;
    Set $v.previous\_current\_balance := i$;
5. **end for**
6. **for** $j$ in $[1, \ldots, N]$ **do**
    **for** $k$ in $[1, \ldots, |V'|]$ **do**
        Set new variable $current\_node := V'[k]$;
        **if** $current\_node.previous\_current\_balance$ is 0 **then**
            Continue;
        **end if**
        Set new variable *opponent* to randomly sampled node from the neighbors of $current\_node$;
        **if** $opponent.previous\_current\_balance$ is 0 **then**
            Continue;
        **end if**
        Set new variable $decision_a$ to the decision (i.e., *stay silent* or *betray*) made by $current\_node$;
        Set new variable $decision_b$ to the decision (i.e., *stay silent* or *betray*) made by $opponent$;
        **if** $decision_a$ is *stay silent* AND $decision_b$ is *betray* **then**
            Set transaction amount $t := min(|P[stay\ silent, betray]|, current\_node.previous\_current\_balance$;
            $current\_node.current\_balance := current\_node.previous\_current\_balance - t$;
            $opponent.current\_balance := opponent.previous\_current\_balance + t$;
        **else if** $decision_b$ is *stay silent* AND $decision_a$ is *betray* **then**
            Set transaction amount $t := min(|P[stay\ silent, betray]|, opponent.previous\_current\_balance$;
            $opponent.current\_balance := opponent.previous\_current\_balance - t$;
            $current\_node.current\_balance := current\_node.previous\_current\_balance + t$;
        **else if** $decision_b$ is *stay silent* AND $decision_b$ is *stay silent* **then**
            Set transaction amount $t := P[stay\ silent, stay\ silent]$;
            **if** $is\_bank\_infinite$ is *True* OR $b \geq 2t$ **then**
                $opponent.current\_balance := opponent.previous\_current\_balance + t$;
                $current\_node.current\_balance := current\_node.previous\_current\_balance + t$;
                **if** $is\_bank\_infinite$ is *False* **then**
                    $b := b - 2t$; # *b cannot be negative because we verified earlier that* $b \geq 2t$
                **end if**
            **end if**
        **else if** $decision_b$ is *betray* AND $decision_b$ is *betray* **then**
            Set $current\_node$ transaction amount $t_1 := min(|P[betray, betray]|, current\_node.previous\_current\_balance)$;
            Set $opponent$ transaction amount $t_2 := min(|P[betray, betray]|, opponent.previous\_current\_balance)$; # *technically, $t_2$ should equal $t_1$ for the symmetric (zero-sum) Prisoner's Dilemma, but this algorithm would work even if this were not the case*
            $current\_node.current\_balance := current\_node.previous\_current\_balance + t_1$;
            $opponent.current\_balance := opponent.previous\_current\_balance - t_2$;
            **if** $is\_bank\_infinite$ is *False* **then**
                $b := b + t_1 + t_2$;
            **end if**
        **end if**
    **end for**
    Compute Gini Coefficient $\mathcal{G}$ using the $current\_balance$ of all nodes in $V'$ using Equation 1;
    Append $\mathcal{G}$ to $L$;
    Break from outer loop if, for all nodes in $V'$, $previous\_current\_balance$ equals $current\_balance$;
    $\forall$ nodes $v \in V'$, set $v.previous\_current\_balance := v.current\_balance$
7. **end for**
8. Output $L$

---

then have 6 units. At this point, A has no units left and cannot, by definition, participate in any further transactions throughout the experiment, either in that iteration or in future iterations. By handling corner cases in this manner, we are able to avoid negative current balances.

Similarly, if the bank is constrained (e.g., has balance strictly less than 2), it will limit how much players can receive from the bank. Because of symmetry, we enforce symmetrical outcomes when the bank is involved i.e., if two players receive anything from the bank, it must always be equal for both. For example, if the bank only has balance of 1 left, and two players each cooperate (stay silent) in a transaction, they will not receive anything from the bank, and the bank's balance stays at 1. In fact, unless other players defect and 'give' money to the bank, cooperating players will always be at a disadvantage in such a situation, leading to an interesting range of behaviors, as subsequently shown.

At the conclusion of each iteration, we compute the *Gini Coefficient* using the current balance of each node. The Gini Coefficient measures statistical dispersion representing the degree of inequality (of current balances) in the network. In economics and policy making, it has been extensively applied to computing wealth and income inequality [18], [46]. The Gini Coefficient $G$ is calculated by averaging the absolute difference of all pairs of node balances in the network, and is equivalently expressed using the formula below:

$$G = \frac{\sum_{i=1}^{|V|}(2i - |V| - 1)x_i}{|V| \sum_{i=1}^{|V|} x_i} \quad (1)$$

Here, $G$ is the Gini Coefficient, $|V|$ is the number of nodes (and equivalently, decision-making agents) in the network, and $x_i$ is the current balance of node $i$.

Given a total number of iterations $N$ as input, Algorithm 1 will simulate the model for up to $N$ iterations or when the balances of all nodes converge, which can occasionally occur well before $N$. In our experiments, $N$ is set to 1,000. Since we compute a Gini Coefficient per iteration, the output list $N$ will contain at most $N$ real-valued elements (showing how Gini Coefficients are evolving as the simulation progresses through iterations), although it may contain fewer elements, if convergence occurred before the $N^{th}$ iteration.

## 4 MATERIALS AND METHODS

### 4.1 Agent Models

This article relies on four classic decision-making models from the game theory and Prisoner's Dilemma literature [50]:

1. **Cooperator:** The cooperator agent model always decides to stay silent.
2. **Defector:** The defector agent model always decides to betray.
3. **Tit-for-Tat:** The tit-for-tat agent model is initialized to stay silent, if its opponent has never transacted before (either with it, or any other agent). If the opponent has transacted before with any agent, the tit-for-tat player will play its opponent's previous decision. Note that the tit-for-tat model can make different decisions for different opponents. In recent years, variants of this model have been proposed owing to its success in Axelrod's tournaments (as well as other models that can outperform tit-for-tat in more complex versions of the game), and its importance in cooperation-based models, but we consider the original version only in this paper, with a fuller investigation of the variants left as future work [55], [56], [57].

4. **Random:** As its name suggests, the random agent model randomly decides (with equal probability) whether to stay silent or to betray each time it is probed for a decision.

### 4.2 Networks and Statistics

**Table 1.** Details on networks used for the experiments in this article. As discussed in the main text, all networks except the Bitcoin OTC network were unweighted and undirected to begin with. The Bitcoin OTC network was converted to its unweighted and undirected equivalent, before being applied in the experiments.

| Network Name | Number of Nodes | Number of Edges | Average Degree | Source (link) |
|---|---|---|---|---|
| Facebook Social circles | 4,039 | 88,234 | 21.85 | https://snap.stanford.edu/data/ego-Facebook.html |
| General Relativity and Quantum Cosmology collaboration network | 5,242 | 14,496 | 2.76 | https://snap.stanford.edu/data/ca-GrQc.html |
| Bitcoin OTC trust weighted signed network | 5,881 | 35,592 | 6.05 | https://snap.stanford.edu/data/soc-sign-bitcoin-otc.html |

The networks used for the experimental studies are based on real-world data and are briefly summarized in Table 1. Each of these networks is publicly available and can be downloaded from the webpage indicated in the table, where more details are also provided on the dataset.

**Facebook Social circles:** This dataset consists of Facebook 'circles' (or 'friend lists'). The data is collected from Facebook app users, with each node being a user along with edges to each of their friends.

**General Relativity and Quantum Cosmology collaboration network:** This network consists of collaborations between authors publishing papers in the fields of General Relativity and Quantum Cosmology. Each node in the network represents an author. When a paper is co-authored by two authors, the graph has an undirected edge connecting the two author nodes. When the paper is co-authored by more than two authors, the graph represents this through a completely connected subgraph on all of the papers' authors.

**Bitcoin OTC trust weighted signed network:** This is a trust network of people who use Bitcoin for transactions on the Bitcoin OTC Platform. Each node represents a user and each edge represent a trade that occurred between two users. The network is weighted to keep a record of user reputation. Members of the platform rate other members on a scale of –10 (indicating complete distrust) to +10 (indicating complete trust). Note that, while the Bitcoin network is technically directed and weighted, for the purposes of this work, we only consider the undirected, unweighted equivalent of the Bitcoin network (since our model currently operates at the level of undirected, unweighted networks), and we ignore the trust ratings. Future work could consider the directed, weighted network once Algorithm 1 has been appropriately extended for such networks.

## 4.3 Experimental Setup

We consider two independent sets of experiments in this work. The first set of experiments investigates the effect of changing the proportions of agent decision-making models in a network. The control for this experiment is simply the case where each of the four models described earlier is assigned to 25% of the nodes in the network. The second set of experiments considers the effects when agent assignment is not random but rather, depends on the *degree* of the node. Specific details on each experiment type are provided below. Both experiment types rely on Algorithm 1 for the simulation, with the sole difference between the two experiment types occurring in line 1 itself (assigning each node in *V* to an agent model). In both experiments, we consider two values for the bank initial balance parameter $b$ (an input to Algorithm 1): 0 and 10,000, with *is_bank_infinite* set to *False*, and we also study the case where *is_bank_infinite* is set to *True* (and where $b$ is not used).

### 4.3.1 Experiment 1: Changing agent proportions

In this experiment, we randomly assign different proportions of nodes to the Cooperator, Defector, Tit-for-Tat, and Random agent models, using a random seed. The control group is the 'equal proportion' case where 25% of the nodes are assigned to each agent model. We consider six other experimental groups that will be compared to the control, summarized in Table 2. Specifically, we increase the proportion of one of the Cooperator, Defector or Tit-for-Tat agent models by 12.5% and simultaneously decrease the proportion of another agent model by the same number, while the random agents' proportion is fixed at 25%.

**Table 2.** Experimental groups used in Experiment 1. The 'equal proportion' control group is in the first row.

| Experimental groups (expressed as proportions) | Percentage of Defector (D) Nodes | Percentage of Cooperator (C) Nodes | Percentage of Tit-for-Tat (T) Nodes | Percentage of Random (R) Nodes |
|---|---|---|---|---|
| D:C:T:R=2:2:2:2 | 25% | 25% | 25% | 25% |
| D:C:T:R=3:1:2:2 | 37.5% | 12.5% | 25% | 25% |
| D:C:T:R=3:2:1:2 | 37.5% | 25% | 12.5% | 25% |
| D:C:T:R=2:3:1:2 | 25% | 37.5% | 12.5% | 25% |
| D:C:T:R=1:3:2:2 | 12.5% | 37.5% | 25% | 25% |
| D:C:T:R=2:1:3:2 | 25% | 12.5% | 37.5% | 25% |
| D:C:T:R=1:2:3:2 | 12.5% | 25% | 37.5% | 25% |

### 4.3.2 Experiment 2: Changing agent assignments based on node degree

In this experiment, we first rank nodes by their degree and assign top, middle, and bottom 1/3 nodes to the Cooperator, Defector, and Tit-for-Tat agent models, respectively. Furthermore, to introduce some randomness, from each of the 1/3 node-sets, we randomly select and assign 25% of the nodes (within each set) to the Random agent. In total, there are six experimental groups that we investigate (Table 3).

## 5 RESULTS

### 5.1 Experiment 1

Figure 2 illustrates the results of simulating Algorithm 1 on the Facebook network. We show one experimental group per subplot, and within each subplot, we show results for the three different bank parameter settings, as discussed earlier. The results show that, when the cooperator proportion is increased (top row) at the *expense* of defectors, the Gini Coefficient of the system starts, and continues, to rise (exhibiting runaway inequality) when the bank has infinite capacity. The intuitive reason for this is that the bank can reward cooperating pairs of individuals indefinitely, as long as there are more cooperating pairs than defecting (or tit-for-tat) pairs. In contrast, when all proportions are equal, the bank's initialization or capacity does not seem to matter much. Interestingly, we find that the relationship is not monotonic. An initial bank balance of 0 for the experimental group D:C:T:R=1:3:2:2 (increasing cooperator proportions at the expense of defectors) leads to higher inequality than a bank balance of 10,000; however, when the bank has infinite supply of capital, the inequality is runaway. This observation shows firsthand the importance of modeling strategic transactions as *entire systems*, since the proportions of other strategies, as well as the presence and power of the external authority, play a much greater role than would be understood through studying the interaction in a silo.

**Table 3.** Experimental groups used in Experiment 2. As discussed in the main text, a quarter of nodes from each of the three sets indicated in the columns below are randomly assigned to the Random agent model.

| Experimental group label | Agent model for top 1/3 nodes | Agent model for middle 1/3 nodes | Agent model for bottom 1/3 nodes |
|---|---|---|---|
| D, C, T | Defector | Cooperator | Tit-for-Tat |
| D, T, C | Defector | Tit-for-Tat | Cooperator |
| C, D, T | Cooperator | Defector | Tit-for-Tat |
| C, T, D | Cooperator | Tit-for-Tat | Defector |
| T, C, D | Tit-for-Tat | Cooperator | Defector |
| T, D, C | Tit-for-Tat | Defector | Cooperator |

Runaway inequality is again observed when the tit-for-tat agent's proportion is increased relative to defector (D:C:T:R=1:2:3:2). Since the tit-for-tat starts as a cooperating agent, and remains cooperating unless it senses that it has a defecting agent, it serves as an approximate proxy for cooperators when it outnumbers the other models. However, it also has a more polarizing effect on the inequality than the C:D:T:R=1:3:2:2 setting.

Interestingly, the system seems to be most stable with respect to bank parameters in the D:C:T:R=3:1:2:2 and D:C:T:R=3:2:1:2 experimental settings, and to a lesser degree, in the control group (D:C:T:R=2:2:2:2). For the former, the defectors dominate at the expense of either cooperator or tit-for-tat agents. Similarly, in the cases where the tit-for-tat agent's proportions are increased relative to the cooperator (or vice versa), we also find that the inequality either stabilizes, or trends toward low inequality. When coupled with the early observation about runaway inequality (when cooperators and tit-for-tat agents are increased at the expense of *defectors*), the simulation expresses that incentives to change strategies don't occur in a vacuum either. Both the initial and resulting strategic mixtures have to be considered in tandem. For instance, incentivizing individuals not to defect in the configuration D:C:T:R=3:2:1:2 is unlikely to have much effect if the expected or desired configuration (as a result of the incentives) is D:C:T:R=2:2:2:2. Compared to the control group, the former actually has lower inequality for all bank settings.

Overall, all of the plots also exhibit a local maximum that is approximately equal to the initial balance (100) assigned to each node. The location of this peak may shift as the balance changes, although more experiments would be needed to verify that hypothesis. There is also likely a dependence on the actual payouts in the Prisoner's Dilemma matrix, which currently (in the non zero-sum cases) are on the order of (gaining) 1 and (losing) 2 for both agents' cooperating and defecting, respectively. As we briefly discuss in Section 7, deriving a theoretical relation between these parameters, the initial balance, the bank setting, network structure, and proportions of agents, and the subsequently observed Gini Coefficient over time (i.e., number of iterations) is a valuable avenue for future research.

In a similar vein as Figure 2, the results for the Physics collaboration and Bitcoin OTC networks are shown in Figures 3 and 4, respectively. There is an impressive degree of consistency between the Physics network and Facebook network in particular, although much consistency is also observed between all three networks. In all three cases, runaway inequality is observed for the two cases we noted earlier (increase in proportions of cooperator, and tit-for-tat, agents, at the expense of the defector agent). The results are also stable in that the (significant) presence of random agents does not distort the overall shape or conclusion of the Gini Coefficient distributions.

Structurally, we find the same positively skewed, normal-like distribution (with local maximum, as noted earlier) in the evolution of the Gini Coefficient in all networks, for all experimental groups. One reason why the local maximum should always be borne in mind, including in future work that attempts to replicate these or other simulations, is the importance of running enough iterations in the simulation. If we had run the simulation even for 100-200 iterations, we may only have observed the peak in all cases, leading us to conclude that inequality will always rise, and winner-takes-all phenomena will always occur in scale-free networks, regardless of bank initialization or strategic mixtures. Indeed, in some cases, there is runaway inequality, but in most cases, the coefficient plateaus after the simulation has been run for a larger number of iterations.

## 5.2 Experiment 2

Figure 5 illustrates the results of simulating Algorithm 1 on the Facebook network for Experiment 2. Although there is still some dependence on the network structure, we find a more straightforward trend compared to the previous experiment. Specifically, when the bank balance is set to a fixed number (0 or 10,000), the Gini Coefficient is always found to converge before reaching the final iteration. When the bank's capacity is set to infinite, Gini Coefficient keeps rising. In other words, if nodes do not sort themselves randomly, but consider their degree when making their decision, and frame their decision-making in a similar vein as other nodes that have a similar degree, the inequality of the system will depend more on the bank than the overall network. The results in this experiment may be a better approximation of the real world, due to established homophily effects in social networks [58].

Similarly, the results for the Physics collaboration and Bitcoin OTC networks are shown in Figures 6 and 7, respectively. Once again, we find the former to be more similar to the Facebook network, and the same qualitative conclusions hold. However, when we consider the latter, we find that, in the two cases (top sub-plots in Figure 7) where the defector is assigned to the top 25% of nodes (ranked by degree), all Gini Coefficients trend to a low value, seeming to converge to a value close to 0. This may be a consequence of the (originally) directed and weighted nature of the Bitcoin OTC network. More study is needed before a conclusion can be reached on those two cases.

## 6 DISCUSSION

We begin this section with a summary of core findings from the earlier section:

1. The results of Experiment 1, wherein we vary the proportions of defector, cooperator, random and tit-for-tat agents to nodes without regard for their structural features (such as degree or clustering coefficient), with different bank settings, shows that runaway inequality tends to occur in two situations:

first, when the cooperator proportion is increased at the expense of defectors, and also when the tit-for-tat agent's proportion is increased relative to defectors.

2. The distribution of the Gini Coefficient tends to peak at the iteration count that coincides with the initial balance assigned to each node at the beginning of the experiment (when there is perfect equality). Following this peak, a range of behaviors is observed, with the coefficient typically trending downward. In some cases (such as the ones noted above), runaway inequality is observed with a momentary decline in inequality followed by a monotonic increase.

3. The bank setting clearly plays an important role in some experiments, especially Experiment 1. However, it is not the deciding factor (on whether runaway inequality is the inevitable consequence) in most cases. Indeed, the conclusion drawn from Experiment 1 is that it is the combination of agent proportions and bank setting that leads to runaway inequality.

4. The Facebook and collaboration networks generally tend to behave similarly, and the Bitcoin OTC network sometimes exhibits trends different from the other two in the same experimental setting. We hypothesize that this is due to its originally directed nature, although more experimentation is needed to test the hypothesis.

5. The results of Experiment 2 further indicate the complications that can arise when network structure and a node's structural positioning within the network (in this case, the node's degree) are considered when assigning agents. Once again, we find the Gini Coefficient following the positively skewed distribution (with a local maximum) in most cases. In a few (but consistent) instances, we also observe a monotonic trend, rather than a local maximum.

When comparing the distributional plots across all three networks and both experiments, we find that, in most cases, a normal-like distribution with positive skew (a thick right tail) is observed. Of course, since the x-axis is discrete time (in this case, modeled as iterations), a negative skew around 0 is not observable. Nevertheless, it may have been observable toward the later iterations. The fact that, in so many cases, the Gini Coefficient peaks precipitously, and often independent of the bank setting, before declining somewhat less precipitously, shows why such social networks and other human systems need to be studied in more dynamical ways than offered by the current literature on social networks.

Another interesting contrast of different experimental groups and bank settings that is only partially understood through the various plots, would be to compare the *final* Gini Coefficients, either upon the $1000^{th}$ iteration or upon convergence (whichever occurs first) in a single table. We provide this data, derived from the same experimental data used to plot the figures in Section 5, in Tables 4 and 5 for the two experiments, respectively.

In Table 4, the highest (final) Gini Coefficient tends to occur in the Bitcoin OTC network, with a coefficient of 0.6 observed for unlimited bank capacity in the 1:2:3:3 group. In considering the averages across all experimental groups, we find that, in every case, inequality is highest in each network when the bank has infinite reserves. Although these results provide some evidence that the structure of the network does play a significant role, with a more direct effect observed in Experiment 2 (compared to Experiment 1) due to the manner in which the experimental groups were designed, they also show the effect of external intervention, even in a simple model such as this one.

Indeed, in recent years, there has been some speculation (both in news media [59] and the scholarly literature [17]) about the United States Federal Reserve's role in the escalating rise in Gini Coefficient that we are currently witnessing in the United States, with levels commensurate to the pre-Great Depression era of the 'roaring twenties'. Although this model is too simplistic to model such a complex entity and the impacts of its monetary policies on the US Gini Coefficient[1], it suggests that, in principle, infinite liquidity can have unwanted systemic effects on the inequality of the system as a whole. Note that we are not actively trying to induce inequality or equality in these experiments; the Gini Coefficient is 0 at the start of every experiment, and in many experimental groups and bank setting combinations, the Gini Coefficient starts trending downward, and even stabilizes, after the local maximum is achieved. However, this is almost never the case, with often severe rises in inequality, especially when considering the results of Experiment 2, when the *is bank setting* is set to *True*.

**Table 4.** A summary of Gini Coefficients observed for each experimental group in Experiment 1 either at the 1000$^{th}$ iteration, or upon convergence (whichever occurs first), for the *bank balance=0/bank balance=10,000/is_bank_infinite=True* settings. Averages are reported to three decimal places.

| Experimental group label (D:C:T:R=) | Facebook | Physics | Bitcoin OTC |
|---|---|---|---|
| 2:2:2:2 | 0.17/ 0.18/ 0.12 | 0.18/0.19/0.15 | 0.04/0.07/0.19 |
| 1:3:2:2 | 0.29/ 0.16/ 0.44 | 0.33/0.35/0.60 | 0.57/0.38/0.58 |
| 1:2:3:2 | 0.13/ 0.16/ 0.52 | 0.34/0.35/0.59 | 0.57/0.56/0.6 |
| 3:1:2:2 | 0.09/ 0.08/ 0.07 | 0.11/0.11/0.22 | 0.09/0.09/0.11 |
| 3:2:1:2 | 0.08/ 0.08/ 0.07 | 0.11/0.28/0.23 | 0.17/0.09/0.21 |
| 2:1:3:2 | 0.35/ 0.17/ 0.12 | 0.24/0.19/0.15 | 0.07/0.07/0.11 |
| 2:3:1:2 | 0.18/ 0.29/ 0.13 | 0.18/0.17/0.15 | 0.04/ 0.04/0.1 |
| AVERAGE | 0.184/ 0.16/ 0.21 | 0.213/ 0.234/ 0.299 | 0.221/ 0.186/ 0.271 |

**Table 5.** A summary of Gini Coefficients observed for each experimental group in Experiment 2 either at the 1000$^{th}$ iteration, or upon convergence, for the *bank balance=0/bank balance=10,000/is bank infinite=True* settings. Averages are reported to three decimal places.

| Experimental group label | Facebook | Physics | Bitcoin OTC |
|---|---|---|---|
| D, C, T | 0.14 / 0.15 / 0.41 | 0.23/0.2/0.43 | 0.04/0.04/0.04 |
| D, T, C | 0.14/ 0.15 / 0.4 | 0.20/0.2/0.54 | 0.04/0.04/0.04 |
| C, D, T | 0.22/ 0.21/ 0.53 | 0.36/0.24/0.65 | 0.4/0.4/0.7 |
| C, T, D | 0.45/ 0.38/ 0.6 | 0.54/0.54/0.7 | 0.38/0.39/0.7 |
| T, C, D | 0.47/ 0.36/ 0.64 | 0.51/0.54/0.7 | 0.4/0.4/0.7 |
| T, D, C | 0.32/0.29/ 0.53 | 0.24/0.27/0.48 | 0.32/0.29/0.53 |
| AVERAGE | 0.29/ 0.257/ 0.518 | 0.347/ 0.332/ 0.583 | 0.263/ 0.26/ 0.452 |

Despite the presence of random elements, we believe that these consistencies illustrate the implicit, or even explicit, role played by the network structure in a larger transactional model. Historically, agents were believed to be rational and independently making decisions e.g., about the expected valuation of a company, or the expected price of an item in an auction [60]. Although much work in behavioral economics has disputed some of these assumptions, the impact of the network structure in which transactions are often conducted, and the presence of external authorities, have both not been empirically and contemporaneously addressed using real networks. Considering that scale-free networks are almost universal, we hope that these findings help spur more research on these systemic effects.

---

[1] The last World Bank estimate of the US Gini Index is 41.1 (equivalent to 0.411 in our model), an almost 1.4 increase over the index in just 2010: https://data.worldbank.org/indicator/SI.POV.GINI?locations=US.

## 7 CONCLUSION AND FUTURE WORK

In this article, we proposed and conducted a set of systematic empirical studies that consider principles at the intersection of game theory, agent-based transactions, and network structure, to simulate and quantify how the inequality of a complex system of inter-connected, transacting agents, as measured by the Gini Coefficient, evolves over time. We proposed a model and simulation algorithm for conducting these studies, and used real-world networks for our experiments. Our results not only showed the impact that agent proportions and network structure can have (or not have, in certain experimental settings), but also illustrated how strong the effects of even a simple external authority can potentially be. Many results were also found to be consistent across different networks, which suggests that fundamental mechanisms are at play in the results we observe.

An important avenue for future research is to consider the functions that govern the Gini Coefficient distributions shown as results for the two experiments, and the theoretical derivation of those functions. It may also be valuable to consider applying the Gini Coefficient in dynamical versions of network growth models, where the network's nodes and edges are themselves not static, but allow for incoming nodes and edges with each iteration. In essence, this would require us to model two kinds of dynamic behavior: the transactional behavior that we explored in this article, and the edge-formation behavior that is often used to explain the scale free degree distributions of the kinds of networks employed in this article. The latter tends to draw on more psychological theories of behavior, such as a preference for new nodes to 'attach' to nodes that (already) have relatively high degree. Some work in game theory has attempted to explain edge formation using transactions, but we hypothesize that both mechanisms (preferential attachment and game-theoretic modeling of cooperative-competitive transactions) can together be more fruitful in producing a richer, more accurate and more theoretically satisfying model of interactive human systems. Finally, it would also be valuable to conduct more experiments, both along the lines of Experiment 2, but with alternative or additional structural features such as centrality and clustering coefficient, as well as more complex bank settings and rules.

## CONFLICT OF INTEREST STATEMENT

The authors declare that the research was conducted in the absence of any commercial or financial relationships that could be construed as a potential conflict of interest.

## AUTHOR CONTRIBUTIONS

Mayank Kejriwal contributed to the conceptualization, methodological design, supervision, funding and the writing, and editing, of the original draft.

Yuesheng Luo contributed to the experimental evaluation, data curation, analysis and writing of the original draft.

## FUNDING

This work was funded by the Defense Advanced Research Projects Agency (DARPA), United States of America with award W911NF2020003 under the SAIL-ON program.

## SUPPLEMENTAL DATA

We release the raw data that was used for generating the figures and tables for *Experiments 1, 2* and *Discussion* as supplemental spreadsheets to enable replication.

## DATA AVAILABILITY STATEMENT

The datasets ANALYZED for this study can be found in the Stanford Network Analysis Project (SNAP) repository: https://snap.stanford.edu/index.html.

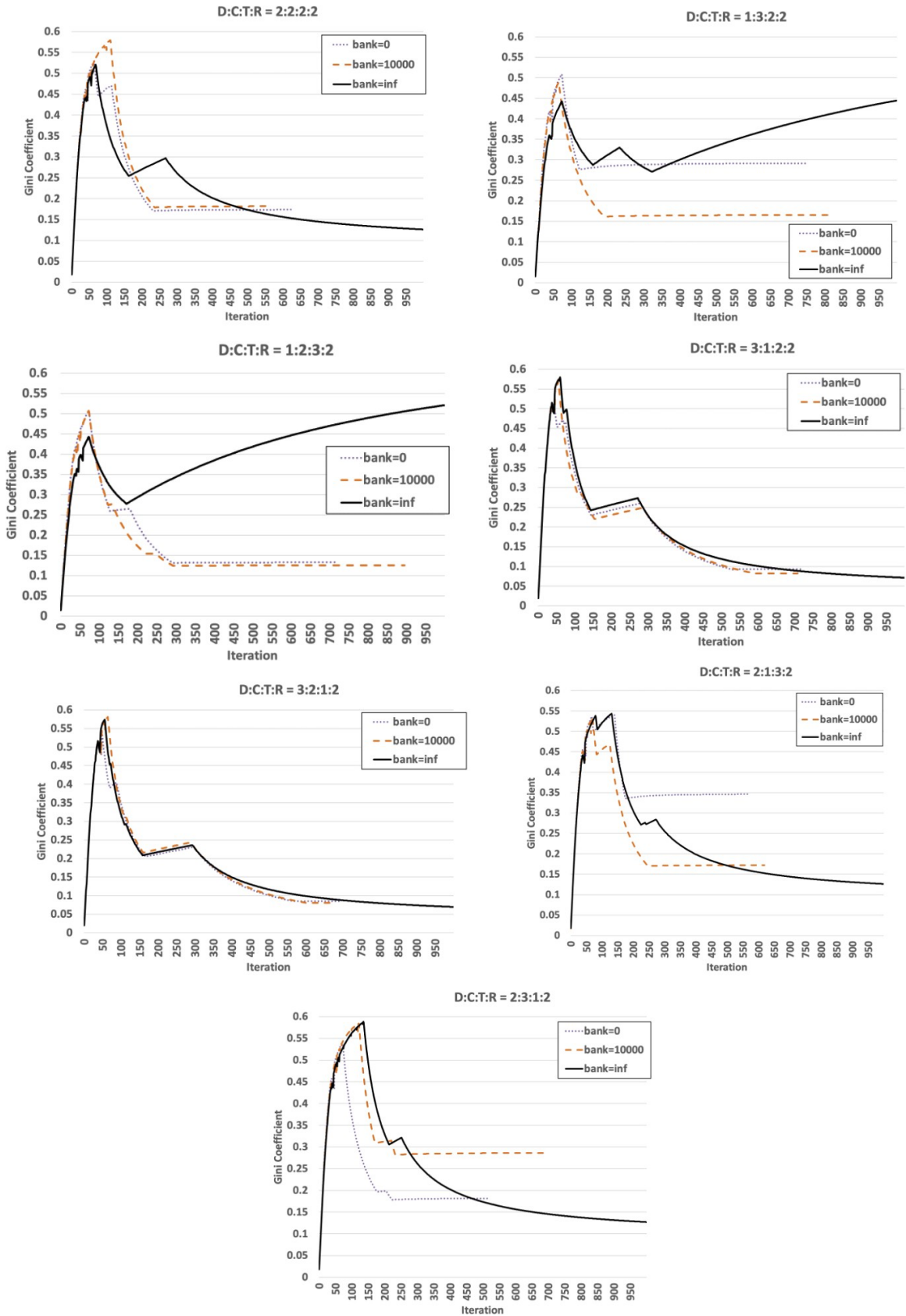

**Figure 2.** Plots of the Gini Coefficient versus number of iterations simulated on the Facebook network, with each subplot illustrating results (using different *bank* parameter values) per experimental group tabulated in Table 2.

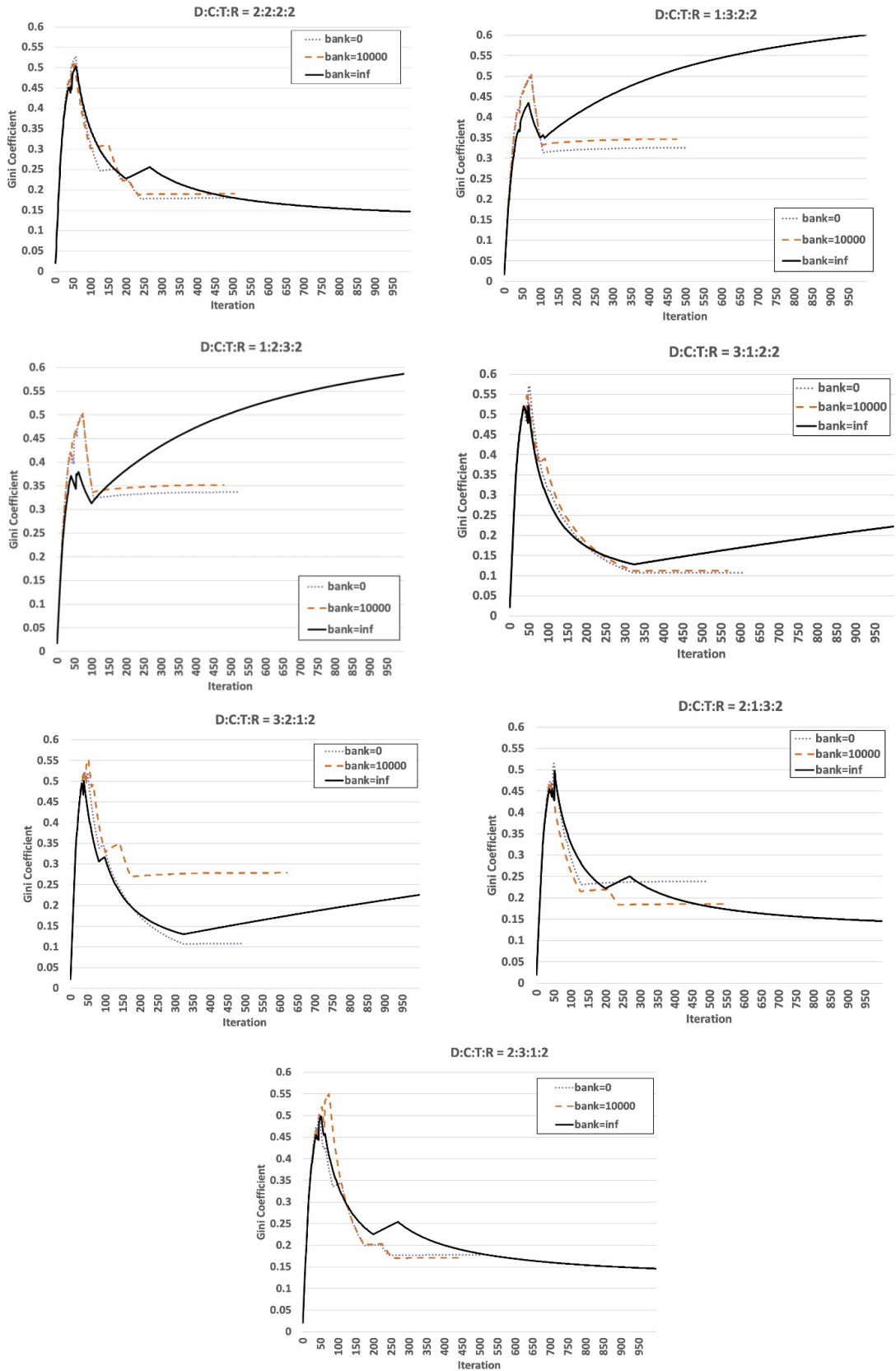

**Figure 3.** Plots of the Gini Coefficient versus number of iterations simulated on the Physics collaboration network, with each subplot illustrating results (using different *bank* parameter values) per experimental group tabulated in Table 2.

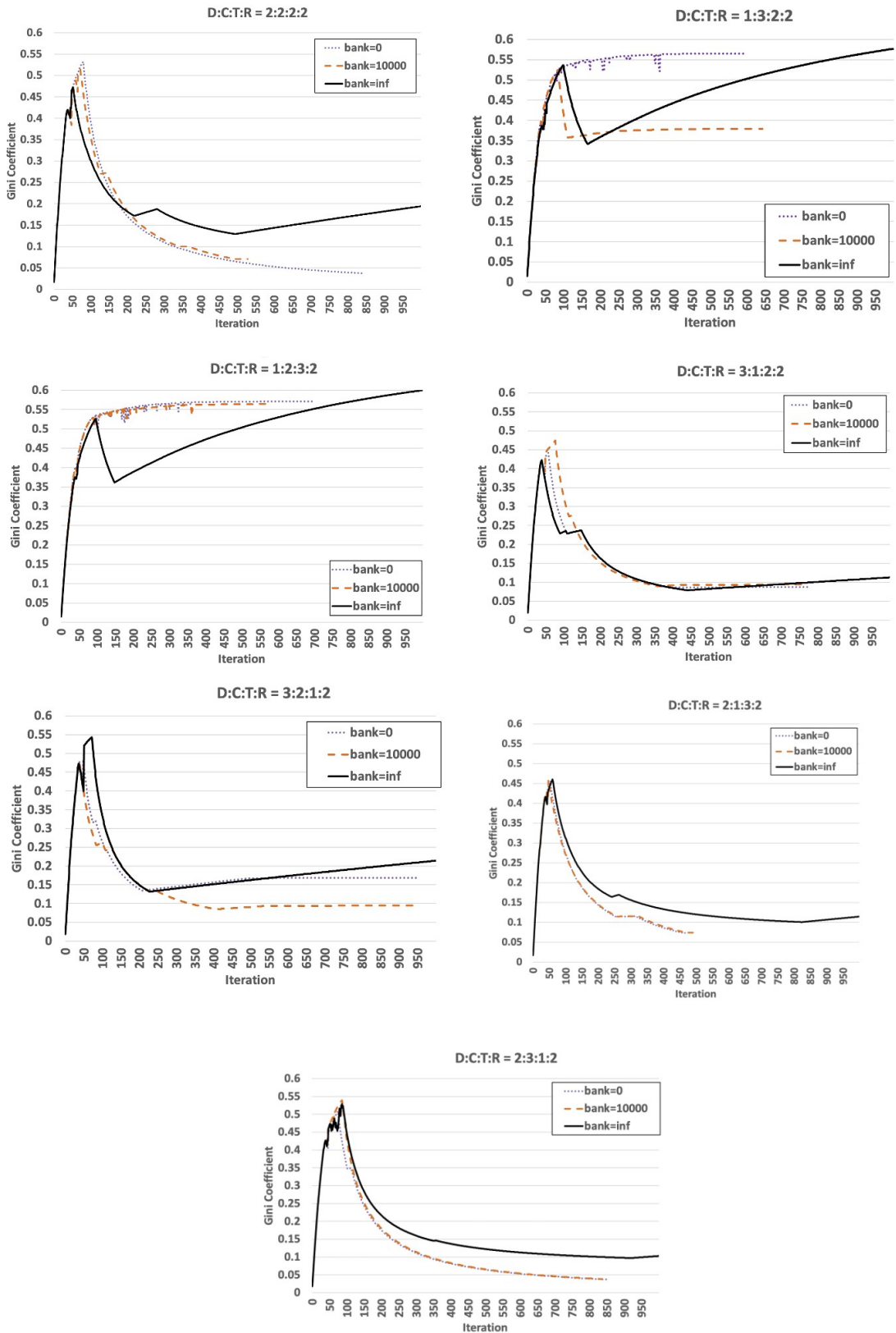

**Figure 4.** Plots of the Gini Coefficient versus number of iterations simulated on the Bitcoin OTC network, with each subplot illustrating results (using different *bank* parameter values) per experimental group tabulated in Table 2.

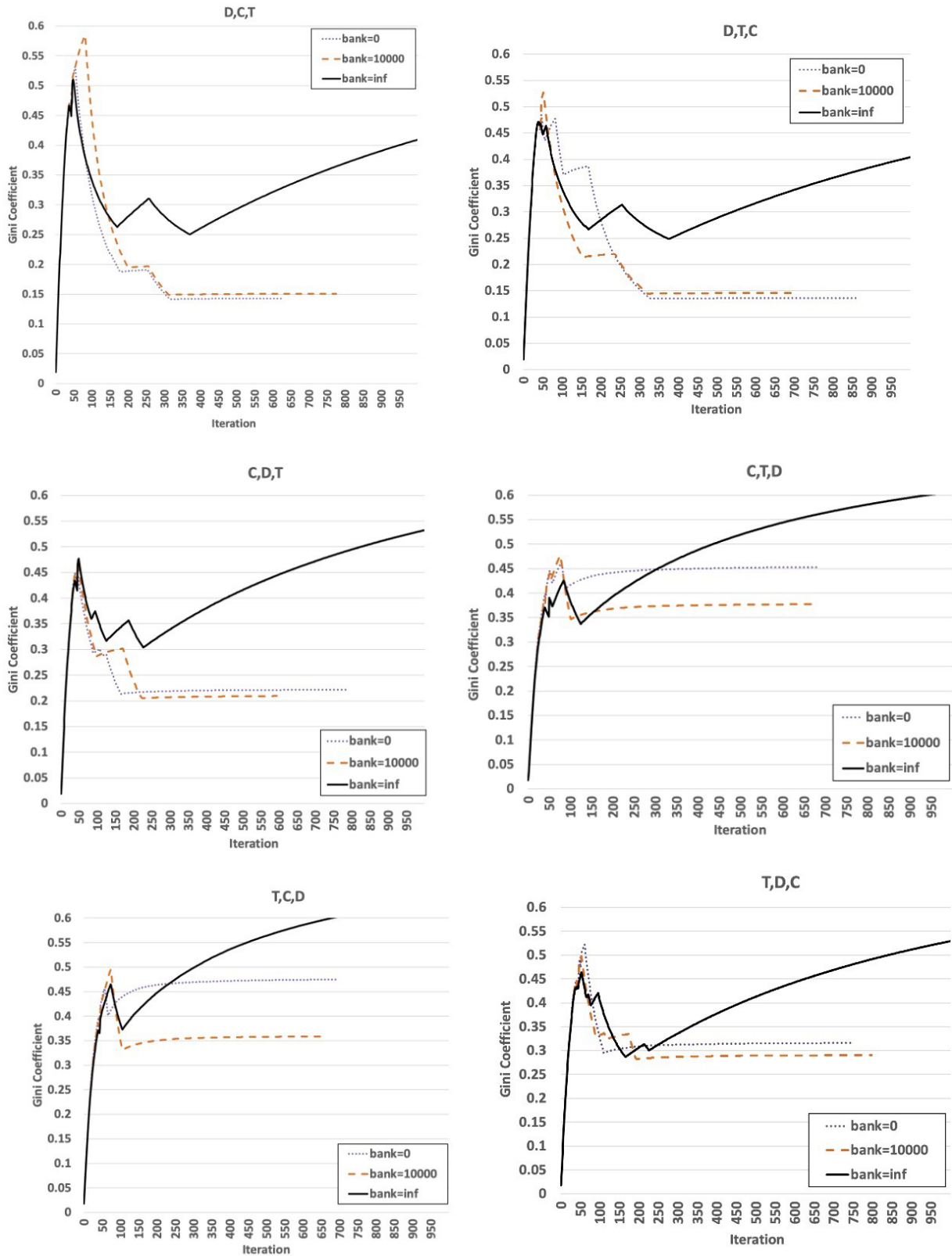

**Figure 5.** Plots of the Gini Coefficient versus number of iterations simulated on the Facebook network, with each subplot illustrating results (using different *bank* parameter values) per experimental group tabulated in Table 3.

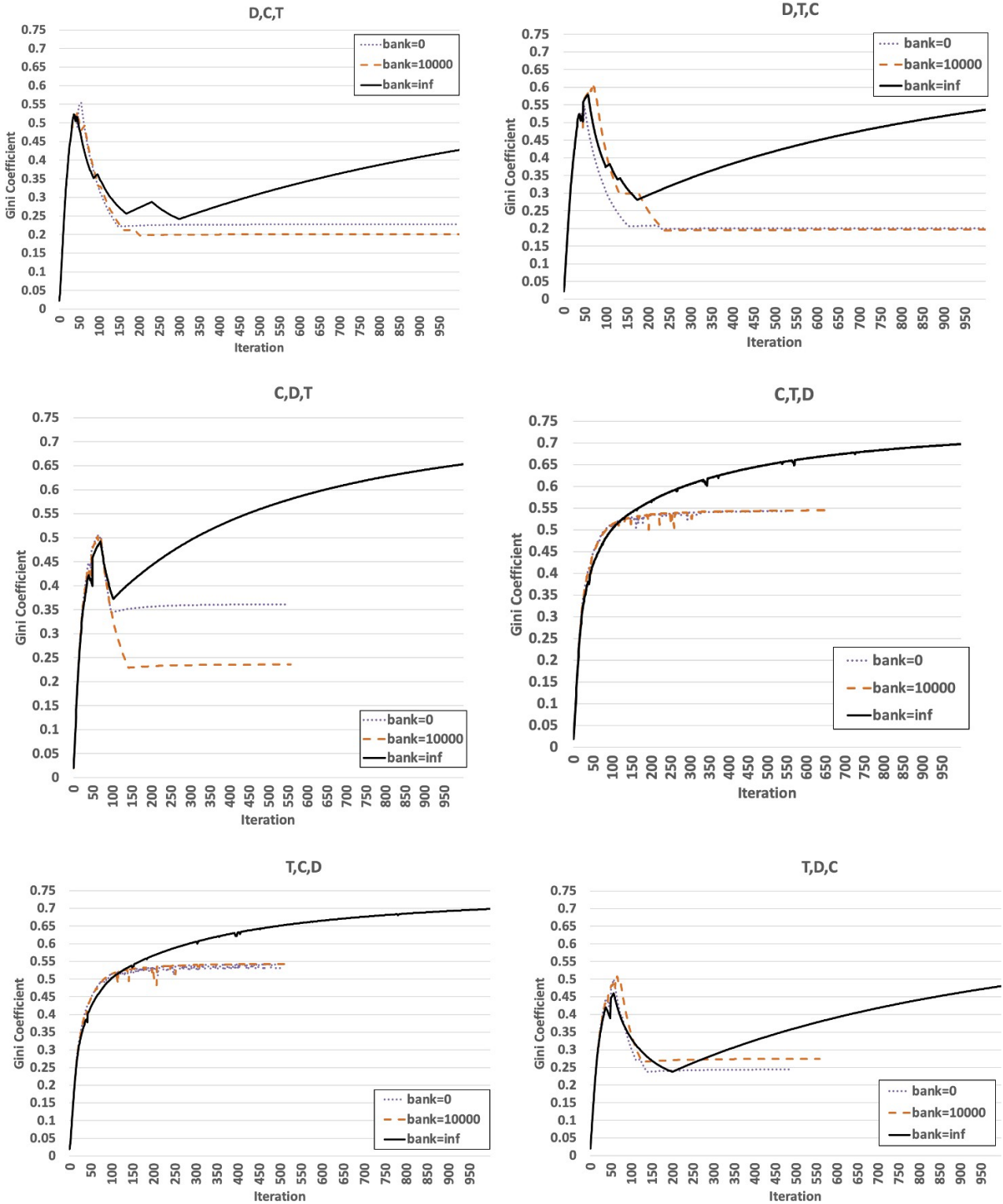

**Figure 6.** Plots of the Gini Coefficient versus number of iterations simulated on the Physics collaboration network, with each subplot illustrating results (using different *bank* parameter values) per experimental group tabulated in Table 3.

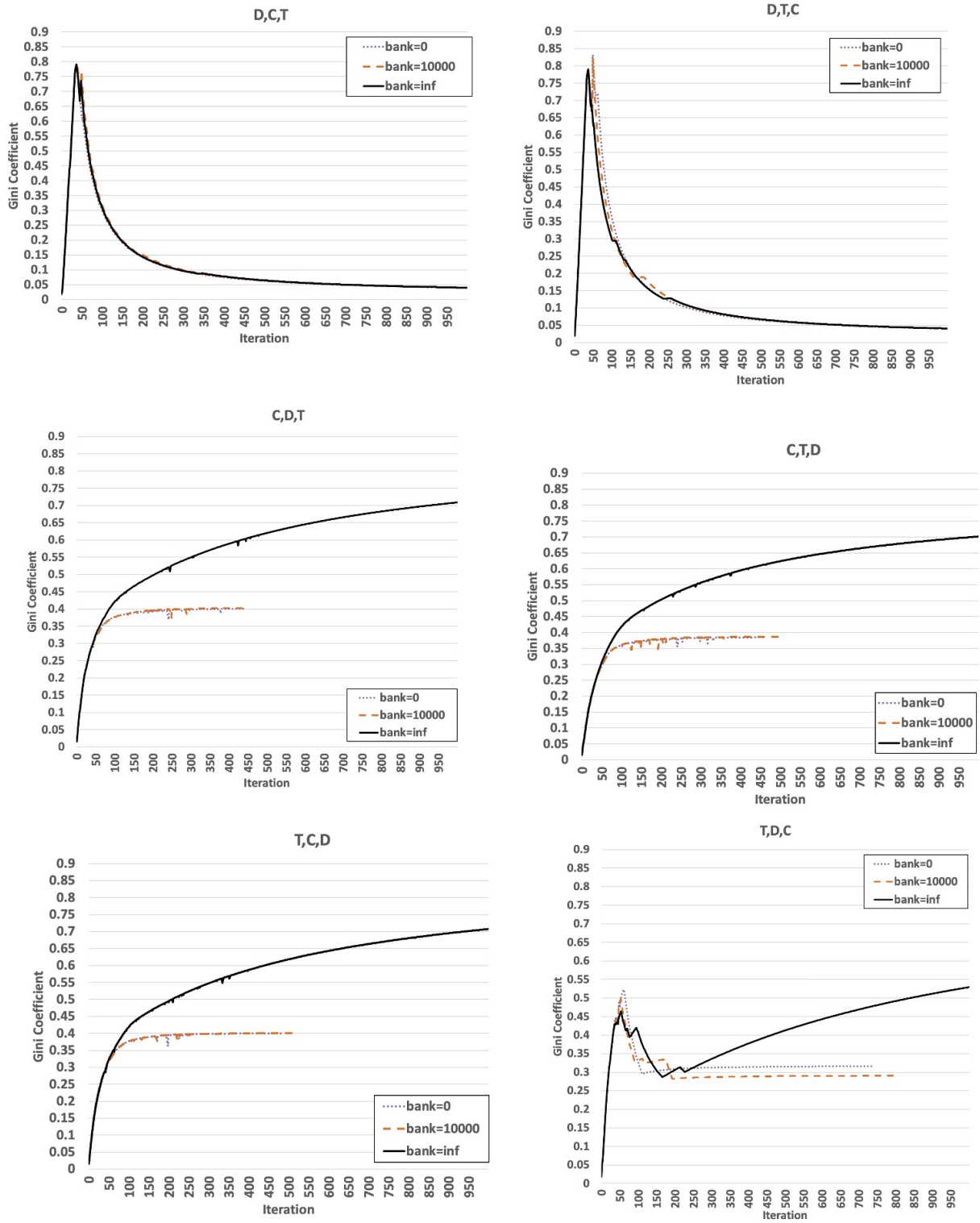

**Figure 7.** Plots of the Gini Coefficient versus number of iterations simulated on the Bitcoin OTC network, with each subplot illustrating results (using different *bank* parameter values) per experimental group tabulated in Table 3.